\documentclass[twocolumn,letterpaper,secnumarabic,amssymb, nobibnotes, aps, prl]{revtex4-1}        
\usepackage{epsfig}
\usepackage{dcolumn}
\usepackage{epstopdf}
\usepackage{graphicx}
\begin{document}

\title{Topological invariants of  electronic currents in magnetic fields}  

 \author{S. Selenu}
\affiliation{}

\begin{abstract}
\noindent  In this article it  is reported a formulation of the solenoidal nature  of quantum electronic currents at the nanoscale whose divergence is expressed as the coupling of a magnetic field, interacting with a quantum body, and a weighted Cern invariant vector making then  a direct topological interpretation of  this quantum magnetic phenomenon. Also a Fourier analysis of the signaling of electronic waves is reported in an ab initio formalism from first pinciples\cite{Martin}.
\end{abstract}

\date{\today} 
\maketitle

 \section{Introduction}
This article may be considered  as a first attempt on regards  the understanding of  electronic current motion  of a quantum system interacting with magnetic fields and is physical properties of being solenoidal as it is well reported by experiments in physical labs\cite{1,2,3}. The reaching of  formulas making us able understanding the very nature of quantum magnetic electronic currents  allows then to perform their  engeneering, lead by the knowledge of their  solenoidal  properties, either at the nano scale or microscopic scale via their control by magnetic fields. It will be shown also how latter properties are related to the very nature of electronic waves nowaday accessible on physical labs them  related to a topological invariant called Berry curvature\cite{Berry0,Berry05} whose physical meaning as long  being evaded since the discovery of the linear connection by M. Berry\cite{Berry1}. It will be shown either the appearence of a Cern invariant vector here thought of being as  the  counterpart of any  measurable classical vector field should we take into account during the engeneering process  of electronic current control across quantum matter. Moreover appears useful to underline that the model of the topological current density is associated to a well measurable vector field  in strict analogy to the classical formulation of  classical electromagnetism\cite{jackson},  where here the leading hypothesis made on classical quantum mechanics is considered, as it is already experimented on electrons at the macroscopic scale\cite{difrazz},  them behaving like waves either at the nanoscale. A formulation of main results therefore obtained will be expressed via an ab initio formalism from first principles in order to obtain  a set of formulas, reached in order, it making us able performing a Fourier analysis of the signal data of electronic waves either on experimental labs. The latter will be derived and reported in the next section where conclusions are left at the end of the article. 
\noindent 
\label{hg}

\section{A topology of electronic currents in magnetic fields}

In this part of the article it will be focused on the developing of a new formulation of magnetism of quantum matter at the nanoscale while concerning the use of magnetic transformations\cite{selenu1} of a quantum eigenstate of matter due to the interaction of the latter with magnetic fields. We may notice how the behaviour of quantum electronic waves due to the scattering of their associated quantum momenta with the classical potential vectors enforces us to take into account a new formalism on the search of solutions of the Shroedinger equation, as found in \cite{selenu1}, with the aim to find the electronic waves across a sample interacting with a magnetic field  it shall be used in order evaluate the quantum electronic current via the transformation of the  linear momentum in the Hilbert space spanned by quantum eigenstates $|\Psi\rangle$. Let us start our evaluation of the expectation value of the linear momentum by considering the unitary transformation acting on eigenstates of the system making changing their wave vector $\bf{k}$ :

\begin{eqnarray}
\label{U1}
U=e^{i{\frac{e}{\hbar c}(\bf{B} \times \bf{r})\cdot {i\nabla_{{\bf{k}}}}}}
\end{eqnarray}

where $e$ is the elementary charge, $c$ the speed of light, and $\bf{B}$ a uniform magnetic field interacting with the quantum body.  On a wave formalism the quantum momentum appears being:

\begin{eqnarray}
\label{U11}
p_n=\langle \Psi_{n,{\bf{k}}-\frac{e}{c}\bf{B \times r}}|\hat{\bf{p}}|\Psi_{n,{\bf{k}}-\frac{e}{c}\bf{B \times r}}\rangle
\end{eqnarray}
it evaluated in the $n$-th state of the electronic wave. The associate electronic current per state is directly calculated as ${\bf{j}}_n=ep_n$, can be studied in an eigenstate reference Hilbert space transforming wave form of the electronic wave signals across the sample body by mean of the quantum magnetic transformations reported in\ref{U1}. In fact, we can directly write:

\begin{eqnarray}
\label{U111}
p_n=\langle \Psi_{n,{\bf{k}}}|\hat{\bf{p}}-\frac{e}{c}{\bf{B \times i\nabla_{\bf{k}}}}|\Psi_{n,{\bf{k}}}\rangle
\end{eqnarray}

showing that with respect to the reference Hilbert space spanned by eigenstates $|\Psi_{n,{\bf{k}}}\rangle$ the macroscopic average of the quantum linear momentum is:

\begin{eqnarray}
\label{U2}
P&&=\sum_n f_n p_n=\sum_n f_n \langle \Psi_{n,{\bf{k}}}|\hat{\bf{p}}-\frac{e}{c}{\bf{B \times i\nabla_{\bf{k}}}}|\Psi_{n,{\bf{k}}}\rangle\\\nonumber
&&=\sum_n f_n \langle \Psi_{n,{\bf{k}}}|\hat{\bf{p}}|\Psi_{n,{\bf{k}}}\rangle -\frac{e}{c}\sum_n f_n \langle\Psi_{n,{\bf{k}}} |{\bf{B \times i\nabla_{\bf{k}}}}|\Psi_{n,{\bf{k}}}\rangle\\\nonumber
&&={\bf{p}}_0 -\frac{e}{c}{\bf{B \times }}[\sum_n f_n \langle\Psi_{n,{\bf{k}}} |i\nabla_{\bf{k}}|\Psi_{n,{\bf{k}}}\rangle]\\\nonumber
&&={\bf{p}}_0 + \frac{e}{c}{\bf{B \times \bar{r}}}={\bf{p}}_0 + \frac{e}{c}{\bf{A}}
\end{eqnarray}

corresponding to  a measurable classical potential vector in classical electromagnetism, where $f_n$ are the occupation numbers of the quantum electronic wave. The associate electric current $J=eP$ carried by the quantal state is then:

\begin{eqnarray}
\label{U2}
J =e{\bf{p}}_0 + \frac{e^2}{c}{\bf{A}}
\end{eqnarray}

At this stage of our modelling it is not yet exploited  the topolical invariance of the quantum divergence  of the electronic current  it shall be shown appearing as the coupling of the Cern invariant with the magnetic vector making us able recognizing the averaged electronic current density being a quantum topological invariant of the system. Let us firstly evaluate  divergence of the electronic current $\nabla \cdot J$  by directly calculating:

\begin{eqnarray}
\label{U22}
\nabla \cdot J&& =e\nabla \cdot{\bf{p}}_0 + \frac{e^2}{Vc}\nabla{\bf{A}}\\\nonumber
&&=\nabla \cdot{\bf{j}}_0 - \frac{e^2}{Vc} B \cdot i[\sum_n f_n \langle i\nabla \Psi_{n,{\bf{k}}} |\times |i\nabla\Psi_{n,{\bf{k}}}\rangle]\\\nonumber 
&&=\nabla \cdot{\bf{j}}_0 -\frac{e^2}{Vc} B \cdot i {\bf{rot } } [\sum_n f_n \langle\Psi_{n,{\bf{k}}} |i\nabla_{\bf{k}}|\Psi_{n,{\bf{k}}}\rangle]\\\nonumber 
&&= \nabla \cdot{\bf{j}}_0 + \frac{e^2}{Vc} B \cdot i {\bf{rot }} [{\bar{\bf{r}}}]
\end{eqnarray}

being ${\bar{\bf{r}}}$ the macroscopic position\cite{seldipolo,dipol} of the quantal body expressed as a function of wave vectors $\bf{k}$.  in view of the fact that  expressing the Berry curvature of the electronic waves, with constant transport numbers\cite{magnetizza}, implie:  

\begin{eqnarray}
\label{Urot}
c&&=  {\bf{rot } } [\sum_n f_n \langle\Psi_{n,{\bf{k}}} |i\nabla_{\bf{k}}|\Psi_{n,{\bf{k}}}\rangle]\\\nonumber
&&= -i [\sum_n f_n \langle i\nabla_{\bf{k}}\Psi_{n,{\bf{k}}}|\times |i\nabla_{\bf{k}}\Psi_{n,{\bf{k}}}\rangle]\\\nonumber
&&=-i [\sum_n f_n \langle i\nabla_{\bf{k}}\Psi_{n,{\bf{k}}}|\Psi_{n,{\bf{k}}}\rangle \times \langle \Psi_{n,{\bf{k}}}|i\nabla_{\bf{k}}\Psi_{n,{\bf{k}}}\rangle]\\\nonumber
&&= i \sum_n f_n [\langle\Psi_{n,{\bf{k}}}| i\nabla_{\bf{k}}\Psi_{n,{\bf{k}}}\rangle \times \langle \Psi_{n,{\bf{k}}}|i\nabla_{\bf{k}}\Psi_{n,{\bf{k}}}\rangle] \langle \Psi_n|\Psi_n\rangle=0\\\nonumber
\end{eqnarray}
by making use of $|\nabla_{\bf{k}} \Psi \rangle = \langle \Psi|\nabla_{\bf{k}}\Psi\rangle |\Psi \rangle$, being also evaluated the vector product of the linear connection with respect  itself being equal to zero. Let us call $C=\frac{1}{V}\int c$ the weigthed Cern invariant, reducing to the Cern invariant for uniformly filled eigenstates. Because of eq.(\ref{Urot}), the Berry curvature or either its weighted counterpart  it is zero for any set of eigenstates, then the volume integral of the divergence of the electronic current is :

\begin{eqnarray}
\label{U222}
&&\frac{1}{V}\int \nabla \cdot{\bf{j}}_0 = 0\\\nonumber
&&\frac{1}{V}\int \nabla \cdot J=  \frac{e^2}{Vc} B \cdot i{\bf{C}}\\\nonumber
&& \cdot {\bf{C}}=0
\end{eqnarray}

where it has been considered a set of electronic waves with constant transport number\cite{magnetizza} on a sample not curring macroscopic electronic currents ${\bf{j}}_0 $ in absence of a magnetic field crossing the  sample itself. Our demonstration of formulas reported in eq. (\ref{U222})  put on a base the topological invarinacy of the quantum divergence  of the electronic current allowing us to notice how a null Cern invariant  gives rise to a null divergence it corresponding to a solenoidal electronic current. The quantum electronic structure\cite{Martin} of the quantum system will then permit  to lead and control the electronic current  on  samples crossed by  external magnetic fields, allowing then to an engeneering of quantum samples at the nanoscale underposed to electronic wave signaling. A useful formula based on Fourier analysis\cite{magnetizza} for the  electronic wave signaling, makes vary the magnitude of the average  electronic currents across the quantum samples,  can be evaluated by an ab initio modelling\cite{magnetizza} as also  reported as a conclusion of our article  in what is the following formulas: 

\begin{eqnarray}
\label{U4}
&&\frac{1}{V}\int \nabla \cdot  J=0\\\nonumber
&&\frac{1}{V}\int J = e {[\sum_{\bf{G}} f_{\bf{G}}|c_{\bf{G}}|^2{\hbar \bf{G}}  ]}+ \frac{e}{c}{\bf{B}} \times [\sum_{\bf{G}}
 f_{\bf{G}} c^*_{\bf{G}}i\nabla c_{\bf{G}} ]\\\nonumber
\end{eqnarray}

 Considering this final  result to put on a new base quantum magnetism by the finding of  formulas involving topological  Cern invariants coupling with the magnetic field as a counterpart to classical magnetism still on use on classical physics, we report conclusions on the  next part of the article.

\section{Conclusions}
   It is shown how   topology of the quantum systems directly influences the magnetic properties of quantum matter where magnetic fields across a sample interacts with the Cern invariant of the quantum system, it not having classical counterpart,  then making this  first principle  ab initio model scheme useful for data analysis in quantum engeneering of electronic currents magnitude of nano and micro magnetic  circuits\cite{jackson}. It have been  reached a calculation of a set of equations can be straigthfowardly used either in the nanoscale modelling of the electronic structure of quantum matter interacting with magnetic fields  or in data fitting of experimental data of electronic currents on physical labs.

\end{document}